\documentclass[aps,prd,floats,preprint,tightenlines,nofootinbib]{revtex4}
\usepackage{rotating}
\usepackage{graphicx}

\begin{document}
\preprint{ \hbox{hep-ph/0506206} } \vspace*{3cm}

\title{Little Higgs Models and Electroweak Measurements}
\author{ Zhenyu Han\footnote{email address:  {\tt zhenyu.han@yale.edu}}
         and  Witold Skiba\footnote{email address:  {\tt witold.skiba@yale.edu}}
       }
\affiliation{  \small \sl  Department of Physics, Yale University,
                New Haven, CT  06520\vspace{2.5cm}
            }

\begin{abstract}
Using effective field theory approach, we study the constraints from
electroweak data on the $SU(5)/SO(5)$, $SU(6)/SP(6)$, $SU(3)\times
U(1)$ little Higgs models and their variations. We construct 
an effective theory valid above the electroweak symmetry breaking 
scale in these models that includes dimension-six operators induced
by integrating out heavy fields. We calculate the constraints on the models
using the results of hep-ph/0412166, where bounds on arbitrary linear
combinations of flavor and CP conserving dimension-six operators are
given. We present the constraints in terms of the bounds on the masses 
of heavy fermions and heavy gauge bosons. The constraints are often stringent,
but  in some regions of the parameter space the constraints are mild enough
and do not imply significant fine-tuning.

\end{abstract}

\maketitle

\newpage

 \section{Introduction}
Little Higgs models
\cite{Arkani-Hamed:2001nc,Arkani-Hamed:2002qx,littlest,su6,Schmaltz:2002wx,
Kaplan:2003uc,Chang:2003un,Skiba:2003yf,Chang:2003zn,Cheng:2003ju,
Cheng:2004yc,Kaplan:2004cr,simplest,Kong:2004cv, Low:2004xc,Thaler:2005en} 
have been proposed to stabilize the electroweak scale in the Standard
Model (SM), see Ref.~\cite{Schmaltz:2005ky} for a review.
In these models, the one-loop quadratically-divergent corrections to
the Higgs mass from the SM particles are canceled by the corrections
from new particles with masses at TeV scale. Eliminating the
one-loop divergences allows the cutoff of the theory to be pushed to
about $4\pi$ TeV, however this could be an optimistic estimate, 
see Ref.~\cite{Chang:2003vs}. The predicted particles are likely to be produced
and observed in future colliders, especially the LHC. The presence
of new particles also creates tension with the electroweak precision
tests (EWPTs). To avoid fine-tuning of more than 10\%, heavy
fermions and gauge bosons with masses less than about  2 TeV and 5
TeV, respectively, should be introduced to cancel the top-loop and
gauge-boson-loop divergences. However, the  EWPTs do not indicate presence of
new particles in a few TeV range if their couplings are generic. 
A successful model has to reconcile the tension between naturalness and the EWPTs.

Constraints on little Higgs models from the EWPTs have been
considered for different models
\cite{Csaki:2002qg,Hewett:2002px,Han:2003wu,
Csaki:2003si,Gregoire:2003kr,Chen:2003fm,Casalbuoni:2003ft,Kilian:2003xt,
Marandella:2005wd}. In this article, we provide a more up-to-date and
extensive analysis employing the effective theory approach we
described in Ref.~\cite{Han:2004az}. In Ref.~\cite{Han:2004az}, we
analyzed all flavor-independent  and CP-conserving dimension-six
operators written in terms of the SM fields that are tightly constrained by EWPTs. 
We calculated the corrections from these operators to the electroweak
precision observables (EWPOs). The result is the $\chi^2$
distribution in terms of the coefficients of these operators. In
an extension of the SM one can integrate out the heavy fields and
obtain the coefficients of the effective operators in terms of
parameters in the model. Substituting the coefficients in the
$\chi^2$ distribution, one immediately obtains global constraints
from all EWPOs. As we will discuss, this procedure fits most little
Higgs models. We will only consider the tree-level diagrams when we
integrate out the heavy fields. Loop diagrams involving heavy fields
are usually suppressed by both the masses of the heavy fields and
the loop factor. Thus, loop corrections from particles with
TeV-scale masses are usually small and do not significantly affect
the constraints we obtain. For the discussion of the fine-tuning problem, we will 
focus on the largest one-loop corrections to the Higgs mass arising from the top quark
and the gauge bosons. Alternative estimates of fine-tuning associated 
with the sensitivity of the Higgs mass to all underlying parameters of little Higgs
 models are presented in Ref.~\cite{Casas:2005ev}.      

 In this article, we focus on the following little Higgs models: the
 $SU(5)/SO(5)$ or the littlest Higgs model \cite{littlest}, the
 $SU(6)/SP(6)$ \cite{su6}
model, and the models with the $SU(3)\times U(1)$ gauge group 
\cite{Kaplan:2003uc,Skiba:2003yf,simplest}, as well as their variations.
For simplicity, from now on we will refer to them as $SU(5)$, $SU(6)$
and $SU(3)$ models respectively, although the first two refer to
their global symmetries and the last one refers to its gauge
symmetry. The $SU(3)$ little Higgs models can have different global
symmetries. In Sec.~\ref{sec:integrate}, we discuss in general what
kind of operators we expect from these models and how to constrain
them. Secs.~\ref{sec:su5}, \ref{sec:su6}, \ref{sec:su3} are devoted
to detailed discussion of each of the three models. We summarize our 
results in Sec.~\ref{sec:conclusions}.

 \section{Integrating out heavy fields}
 \label{sec:integrate}

A complete set of independent dimension-six operators in the SM is
given in Ref.~\cite{Buchmuller:1985jz}. Assuming flavor and CP
conservation, in Ref.~\cite{Han:2004az} we narrowed this set down to
21 operators that are relevant to EWPTs. In a compact notation, the operators
are:
\begin{eqnarray}
   O_h&=&| h^\dagger D_\mu h|^2,\nonumber\\
  O_{hf}^s& =& i (h^\dagger D^\mu h)(\overline{f} \gamma_\mu f) + {\rm h.c.}, \ \ \
  O_{hf}^t = i (h^\dagger \sigma^a D^\mu h)(\overline{f} \gamma_\mu \sigma^a f)+
     {\rm h.c.},\nonumber\\
 O_{ff'}^s&=&\frac{1}{1+\delta_{ff'}} (\overline{f} \gamma^\mu f) (\overline{f'} \gamma_\mu f'), \ \ \
 O_{ff'}^t=\frac{1}{1+\delta_{ff'}} (\overline{f} \gamma^\mu \sigma^a f)  
                 (\overline{f'} \gamma_\mu \sigma^a f'),\label{ops}
\end{eqnarray}
where $h$ is the SM Higgs doublet, $f,f'=q,l,u,d,e$, are the left
and right handed fermions. The operators are understood to be summed
over flavor indices. The superscripts $s$ and $t$ stand for singlet
and triplet $SU(2)$ contractions. For triplet couplings, $f$ has to be a SM
doublet ($q$ or $l$). Note that $O_h$ corresponds to the oblique $T$
parameter \cite{Peskin:1991sw} that breaks the custodial symmetry.
We have omitted two operators from our list.   The omitted operators are denoted
$O_{WB}$ and $O_W$ in Ref.~\cite{Han:2004az}. The former corresponds
to the oblique $S$ parameter and the latter modifies triple
gauge-boson couplings. These two operators are not induced at tree level in the
models analyzed here. Four-fermion operators involving only quark
fields were not included in our list because they are not constrained
as tightly as operators involving some leptons.  

Including operators $O_i$, we can write the effective
Lagrangian as
\begin{equation}
  \mathcal{L}=\mathcal{L}_{SM}+\sum_i a_i O_i.
\end{equation}
In Ref.~\cite{Han:2004az} we calculated the corrections to EWPOs
from the operators $O_i$ assuming arbitrary coefficients $a_i$. The
corrections were combined with known experimental values and the SM
predictions to obtain the total $\chi^2$ distribution
\begin{equation}
  \chi^2=\chi^2(a_i)=\chi^2_{min} + 
                (a_i - \hat{a}_i)  {\mathcal M}_{ij} (a_j-\hat{a}_j),\label{chi2}
\end{equation}
where $\hat{a_i}$ are values of $a_i$ that minimize $\chi^2$. In
Ref.~\cite{Han:2004az}  we calculated the corrections to EWPOs 
to linear order in $a_i$, so $\mathcal{M}$ in Eq.~(\ref{chi2}) is a constant
and positive-definite matrix. In a given model, we integrate out the
heavy fields and obtain the coefficients $a_i$ as functions of the
parameters in the model. Expressing $a_i$ in Eq.~(\ref{chi2}) 
in terms of parameters of a model allows us to immediately obtain
constraints on the model without having to compute EWPOs.

Before we list and analyze in detail the coefficients of
effective operators in the little Higgs models, we briefly
discuss how the effective operators are generated
and what interesting features these models have.

There are three kinds of heavy fields in these models: gauge bosons,
scalars and fermions. We first discuss the effects of the heavy
gauge bosons. The $SU(5)$ and $SU(6)$ models share the same gauge
structure: $[SU(2)\times U(1)]^2$. The gauge group is broken to the
diagonal $SU(2)\times U(1)$ which is identified with the SM gauge
group. Thus half of the gauge bosons get masses and the
other half remain massless until electroweak symmetry breaking takes place. 
The heavy gauge bosons include a triplet $W'$ and a singlet $Z'$, which couple
to both the SM Higgs and fermions. The exchange of $W'$ generates
the triplet coupling operators $O_{hf}^t$, $O_{ff'}^t$, while the
exchange of $Z'$ generates $O_h$, $O_{hf}^s$, $O_{ff'}^s$. The gauge
sector in the $SU(3)$ model is quite different: a gauged
$SU(3)\times U(1)$ is broken to the SM $SU(2)\times U(1)$, leaving 5
heavy gauge bosons. One of them behaves like the $Z'$ and also
induces $O_h$, $O_{hf}^s$ and $O_{ff'}^s$. The others decouple from
the light fields. There are no $SU(2)$ triplet operators generated in
this case.

Now we turn to discussing the scalars in the models. In the low-energy
spectra, the $SU(5)$ and $SU(3)$ models contain one Higgs doublet
and the $SU(6)$ model contains two Higgs doublets. The number of
light Higgs doublets is irrelevant in our analysis since  only the
Higgs vev matters. Besides the doublet, the $SU(5)$ model also
contains a heavy triplet scalar. When integrated out, the triplet
generates the $O_h$ operator. In addition, there are heavy singlets
in all the three models, but integrating them out does not generate
dimension-six operators relevant to EWPTs.

Turning to fermions, heavy fermions are needed to cancel the
quadratic divergence from the top loop. However, there are often multiple
choices that insure the cancelation. In the $SU(5)$ little Higgs model
\cite{littlest}, a pair of vector-like heavy fermions is added to
the SM fields. The right-handed heavy fermion mixes with the
right-handed top quark so that the couplings between the top quark
and the SM gauge bosons are modified. The loop divergences from the
light two generations do not introduce fine-tuning for a cutoff as
low as $4\pi$ TeV because  the corresponding Yukawa couplings are small.
Thus one does not introduce extra fermions to cancel the divergences
for the light two generations. Therefore, fermion couplings in this
model are flavor-dependent. However, since only the top quark mixes
with the heavy fermion and no EWPO involves the top quark in the
final or initial states, this flavor-dependent effect is not
relevant. Of course, if we added fermions that mix with the first 
two generations as well, we could generate
flavor-dependent operators that do affect EWPTs. Such
operators can introduce FCNCs and would be severely
constrained. A detailed analysis for this case is beyond the scope
of this paper. We will assume approximate flavor-independence in our
analysis.

The fermion sector in the $SU(6)$ model is similar to the $SU(5)$
model\footnote{In Ref.~\cite{su6}, the Yukawa structure also
introduces mixing for the bottom quark, but we can make a similar
choice as in the SU(5) model. Sec.~\ref{sec:su6} contains an
example.}, while the $SU(3)$ model merits a few more comments.
Because the gauge group contains an $SU(3)$, every SM fermion
doublet  must be combined with an extra fermion to complete an
$SU(3)$ triplet. Thus unlike the $SU(5)$ or the $SU(6)$ model, heavy
fermions have to be added to all generations. If we assign fermions
the same quantum numbers for the three generations and impose the
constraint of small FCNCs, we will obtain flavor-independent
operators $O_{hq}^{s,t}$ and $O_{hl}^{s,t}$ by integrating out the
heavy fermions. This is a result of mixing between the SM and the 
heavy fermions. This mixing modifies the gauge couplings of the SM fermions.

Besides the operators obtained by integrating out the heavy fields,
there exist dimension-six operators arising from expanding the kinetic term of
the nonlinear sigma field to higher orders. It turns out that the dimension-six
operators obtained this way do not affect EWPTs in the $SU(5)$ and
$SU(3)$ models, while in the $SU(6)$ model, there are contributions
to the $O_h$ operator when $\tan\beta\neq1$, where $\tan\beta$ is
the ratio of the two Higgs vevs.

\section{The $SU(5)/SO(5)$ model \cite{littlest}}
\label{sec:su5}

Detailed description of the $SU(5)$ little Higgs model can be found
elsewhere. We will only specify the necessary conventions and notation.
Throughout this paper, we use $h=(h^+\mbox{ }
h^0)^T$ to denote the Higgs doublet, $v$ the vev of the Higgs, and
$g$ and $g'$ the gauge coupling constants of the SM. The littlest Higgs
model is based on a nonlinear sigma model with an $SU(5)$ global symmetry
spontaneously broken to its $SO(5)$ subgroup. The $SU(5)$ breaking direction
is given by the vev of the  $\Sigma$ field
\begin{equation}
  \Sigma_0\equiv\langle\Sigma\rangle=
    \left(\begin{array}{ccc}&&\textbf{1}_2\\&1&\\\textbf{1}_2&&\end{array}\right).
\end{equation}
$\Sigma$ can be parameterized around its vev  as
\begin{equation}
  \Sigma=e^{2i\Pi/ F}\Sigma_0,
\end{equation}
where the Goldstone boson matrix $\Pi$ is defined as
\begin{equation}
  \Pi=\left(\begin{array}{ccc}0&\frac{\tilde{h}}{\sqrt{2}}&\phi^\dag\\
         \frac{\tilde{h}^\dag}{\sqrt{2}}&0&\frac{\tilde{h}^T}{\sqrt{2}}\\
         \phi&\frac{\tilde{h}^*}{\sqrt{2}}&0\end{array}\right).
\end{equation}
In the equation above, $\tilde{h}=i\sigma^2 h^*$, and $\phi$ is a
two by two symmetric matrix that represents a  scalar triplet with
mass $M_\phi$ of order $F$.  We have omitted the
fields that are eaten when the gauge group is broken to the SM gauge
group. The kinetic term of the nonlinear sigma model is given by
\begin{equation}
\frac{F^2}{8}\mbox{Tr}[D_\mu \Sigma
D^\mu\Sigma^\dag],\label{kinetic}
\end{equation}
where the covariant derivative is defined as
\begin{equation}
   D_\mu\Sigma=\partial_\mu \Sigma - i\sum_{j=1}^2[g_jW_j^a(Q_j^a\Sigma+
            \Sigma Q_j^{aT})+g_j'B_j(Y_j\Sigma+\Sigma Y_j^T)]\label{derivative}.
\end{equation}
$Q_j^a$ and $Y_j$ are the generators of the $[SU(2)\times U(1)]^2$
gauge group, defined in the same way as in Ref.~\cite{littlest}.
$W_j^a$ and $B_j$ are the corresponding gauge bosons. The mass
eigenstates of the gauge bosons are 
\begin{eqnarray}
   W&=&sW_1+cW_2, \quad W'=-cW_1+sW_2, \nonumber\\
   B&=&s'B_1+c'B_2,\quad Z'=-c'B_1+s'B_2,\label{gaugemixing}
\end{eqnarray}
where
\begin{equation}
s,c=\frac{g_2,g_1}{\sqrt{g_1^2+g_2^2}},\quad
s',c'=\frac{g_2',g_1'}{\sqrt{g_1'^2+g_2'^2}}.\label{mixingangles}
\end{equation}
In Eq.~(\ref{gaugemixing}), $W$ and $B$ are the SM gauge bosons,
and $W'$ and $Z'$ are heavy gauge bosons with masses
\begin{equation}
M_{W'}=\frac{gF}{2sc},\quad
M_{Z'}=\frac{g'F}{\sqrt{20}s'c'},\label{gaugebosonmass}
\end{equation}
where $g=g_1 s=g_2 c$ and
$g'=g_1' s'=g_2' c'$

Most corrections to EWPOs come from the exchanges of the $W'$ or $Z'$
bosons. Another correction comes from integrating out the triplet
field $\phi$ which couples to $h$ as
\begin{equation}
-i\lambda (h^T\phi^\dag h-h^\dag\phi h^*),
\end{equation}
where $\lambda$ is a dimensionful coupling of order $F$. The coefficient $\lambda$
is not determined in the low-energy theory since it arises from
quadratically-divergent contributions. When integrated out, the triplet generates
both dimension-four and dimension-six operators. The dimension-four operator
is the $|h|^4$ term contributing to the Higgs potential. In order to
insure that the Higgs potential is bounded from below and generates
the correct vev for EWSB, the following relation has to be satisfied
\cite{Han:2003wu}
\begin{equation}
  \frac{\lambda^2 F^2}{M_{\phi}^4}<\frac14.\label{lambdarelation}
\end{equation}
Besides the dimension-four operator, the triplet also generates the
dimension-six $O_h$ with the coefficient $2\lambda^2/M_{\phi}^4$.
We will treat $\lambda^2/M_{\phi}^4$ as a free parameter that is
subject to the constraint (\ref{lambdarelation}). Note that if
we did not integrate out the triplet, after EWSB the triplet would
obtain a vev that breaks the custodial symmetry:
\begin{equation}
v'=\langle\phi\rangle=\frac{\lambda v^2}{2M_\phi^2}.
\end{equation}
This is another way to understand the contribution to $O_h$.

We assume that fermions are all singlets under the second $SU(2)$ as
in Ref.~\cite{littlest}, but allow them to be charged under both
$U(1)$'s. The model includes a pair of vector-like heavy fermions to
cancel the top loop contribution to the Higgs mass. As explained in
Sec.~\ref{sec:integrate}, the heavy fermions do not affect EWPOs.

Integrating out the heavy fields $W'$, $Z'$ and $\phi$, we obtain:
\begin{eqnarray}
a_h&=&-\frac{5(c'^2-s'^2)^2}{2 F^2}+\frac{2\lambda^2}{M_\phi^4},\nonumber\\
a_{hq}^t&=&a_{hl}^t=-\frac{(c^2-s^2)c^2}{2 F^2},\nonumber\\
a_{hf}^s&=&\frac{5s'c'(c'^2-s'^2)}{F^2}\left(Y_2^f\frac{s'}{c'}-Y_1^f\frac{c'}{s'}\right),\nonumber\\
a_{lq}^t&=&a_{ll}^t=-\frac{c^4}{F^2},\nonumber\\
a_{ff'}^s&=&-\frac{20s'^2c'^2}{F^2}\left(Y_2^f\frac{s'}{c'}-Y_1^f\frac{c'}{s'}\right)
\left(Y_2^{f'}\frac{s'}{c'}-Y_1^{f'}\frac{c'}{s'}\right),\label{littlestops}
\end{eqnarray}
where $Y_1^f$ and $Y_2^f$ are the charges of fermion $f$ under the
two $U(1)$'s.  $Y_1^f$ and $Y_2^f$ are assumed to be generation independent. 
The SM hypercharge is $Y=Y_1+Y_2$. If
the $U(1)$ charge assignment is given, we can substitute the coefficients
$a_i$ into Eq.~(\ref{chi2}) and obtain $\chi^2$ as a function of $f$,
$c$, $c'$ and $\lambda^2/M_\phi^4$.

As we will see shortly, the coefficients $a_i$ can  put
tight constraints on $F$. If $F$ is too large, we will reintroduce
fine-tuning to the theory. Therefore it is interesting to consider
how to choose the parameters in Eqs.~(\ref{littlestops}) to suppress
$a_i$. It is easy to see that if $c\ll1$ and $s'\approx c'$, the
coefficients $a_h$, $a_{hf}^s$, $a_{hf}^t$ and $a_{ff'}^t$ are all
suppressed. If we further assume that the fermions are charged
equally under the two $U(1)$'s ($Y_1^f=Y_2^f$), the coefficients
$a_{ff'}^s$ also vanish. However, it is impossible to render the two
$U(1)$ charges equal unless one allows the $U(1)$'s to be outside of
the global $SU(5)$. Changing $Y_1\rightarrow Y_1 +bI$, and
$Y_2\rightarrow Y_2-bI$ and setting $b=-1/80$ \cite{Csaki:2003si},
we can obtain equal charges\footnote{In this case, the coefficients
listed in Eqs.~(\ref{littlestops}) are modified because of the
change in the $Z'$ mass. The coefficients $a_{hf}^t$ and $a_{ff'}^t$
do not change while $a_{hf}^s$, $a_{ff'}^s$ and the term multiplying
$(c'^2-s'^2)^2$ in $a_h$ are rescaled by a factor of
$1/(1+100b^2)$.}. These parameter choices are very much like the
``near-oblique'' limit for the $SU(6)$ model discussed in
Ref.~\cite{Gregoire:2003kr}. In this limit, the only significant
corrections come from the $\lambda^2/M_\phi^4$ term in $a_h$.
In addition, $c$ can not be arbitrarily small  and has to be greater than 
$g/4\pi$, otherwise the second $SU(2)$ will be strongly coupled.

\begin{figure}
\begin{center}
\includegraphics[width=\textwidth]{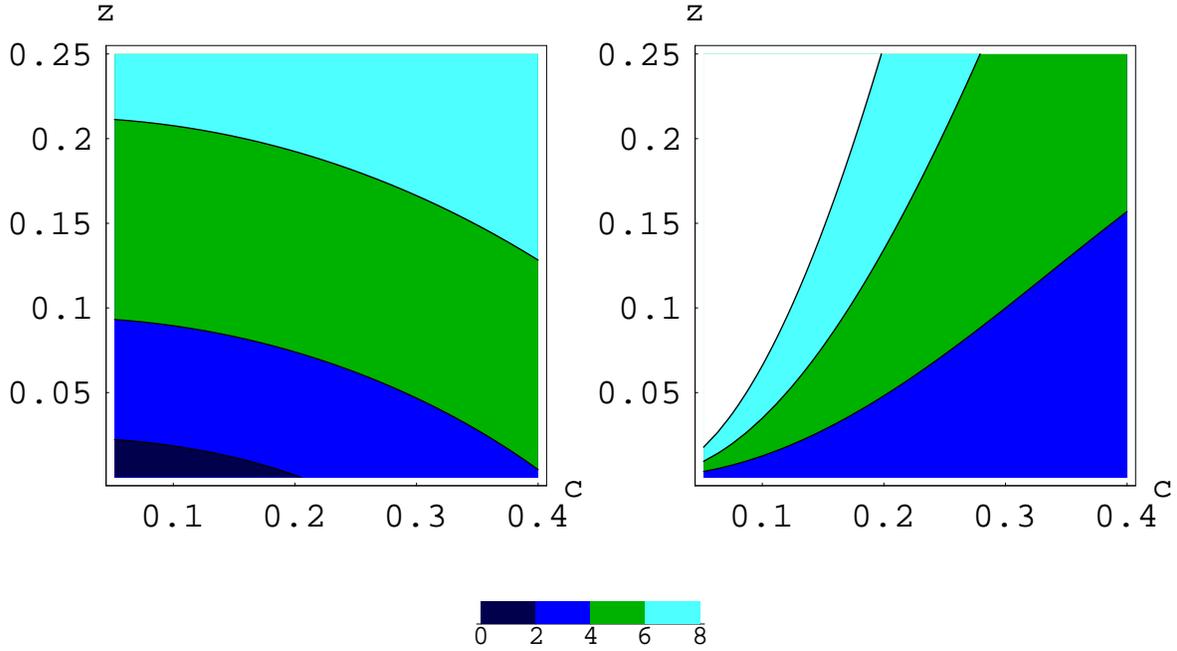}
\end{center}
\caption{95\% CL lower bounds in TeV  on $M_{t'}$ (left) and
$M_{W'}$ (right) in the $SU(5)$ model 
as functions of $c$ and $z \equiv \lambda^2 F^2/M_\phi^4$ for
$Y_1^f=Y_2^f$ and $s'=c'$. In the plots, $c\in[g/4\pi,0.4]$ and
$z\in[0,1/4]$.}
\label{fig:littlest1}
\end{figure}

To make the constraints and thus the associated fine-tuning more
transparent, we trade $F$ for the heavy top mass $M_{t'}$ and the
heavy triplet gauge boson mass $M_{W'}$. For the Higgs mass of order
200~GeV, 10\% fine-tuning corresponds to $M_{t'}\sim2$ TeV or
$M_{W'}\sim 6$ TeV. $M_{W'}$ is given in
Eq.~(\ref{gaugebosonmass}). As discussed in Ref.~\cite{littlest},
$M_{t'}\geq\sqrt{2}\lambda_t F$, where $\lambda_t\sim 1$ is the top
Yukawa coupling. For simplicity, we set $M_{t'}=\sqrt{2} F$.
Fig.~\ref{fig:littlest1} shows the bounds on $M_{t'}$ and $M_{W'}$
around the ``near-oblique" limit as functions of $c$ and
$\lambda^2 F^2/M_\phi^4$. We have set $s'=c'$ and $Y_1^f=Y_2^f$ in
the two plots. As expected, the bounds on $M_{t'}$ are loose near
$c=g/4\pi$ and $\lambda^2f^2/M_\phi^4=0$. On the other hand, because
$M_{W'}\propto1/sc$, the bounds on $M_{W'}$ are quite tight near
$c=g/4\pi$, introducing more fine-tuning than that from the top
sector. Nevertheless, there clearly exists a region with less than
10\% fine-tuning. It would be desirable if the charge assignment and
the parameter space limit could come naturally from a UV extension of
the model.

\begin{figure}
\begin{center}
\includegraphics[width=\textwidth]{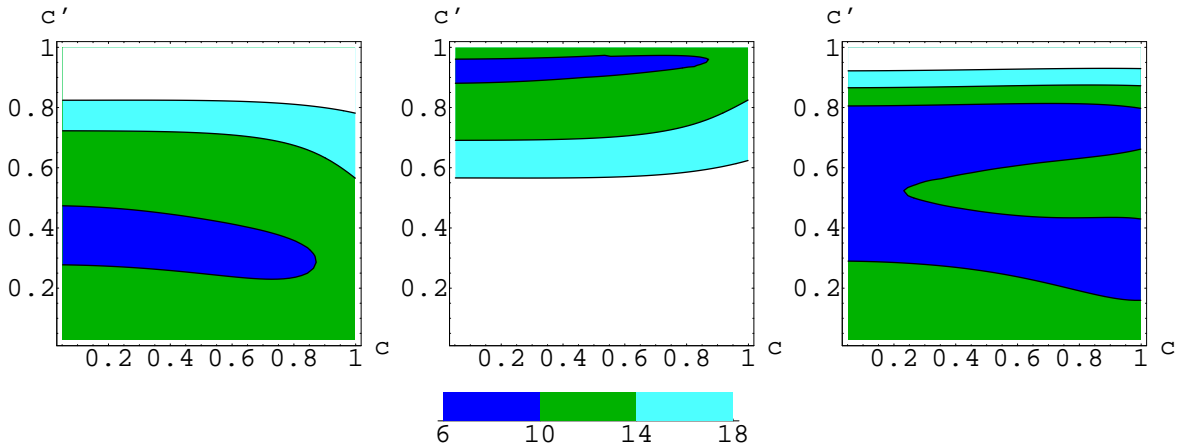}
\end{center}
\caption{ Lower bounds on $M_{t'}$ in TeV in the $SU(5)$ model as a function of
 $c$ and $c'$ at 95\% CL. Left: $U(1)$ charge assignment $Y_1^f=Y^f$,
$Y_2^f=0$; middle: $Y_1^f=0$, $Y_2^f=Y^f$; right: $Y_1^f=Y^f$,
$Y_2^f=0$ but LEP2 data is excluded. In the plots, $c\in
[g/4\pi,\sqrt{1-(g/4\pi)^2}]$ and
$c'\in[g'/4\pi,\sqrt{1-(g'/4\pi)^2}].$} \label{fig:littlest2}
\end{figure}

Another way to suppress the coefficients in Eqs.~(\ref{littlestops})
is by taking $c,c'\ll1$ and assuming the fermions are charged under only the
first $U(1)$ ($Y_1^f=Y^f$, $Y_2^f=0$). In this limit, all
coefficients except $a_h$ are suppressed. It turns out that $a_h$
alone can still put tight constraints on $F$, as can be seen from
Fig.~\ref{fig:littlest2}. In Fig.~\ref{fig:littlest2}, we plot  the
95\% confidence level (CL) bounds on $M_{t'}$ for two fermion charge assignments, as a
function of $c,c'$. The left (middle) plot corresponds to all
fermions charged under only the first (second) $U(1)$. In the
plots, we allow both $M_{t'}$ and $\lambda^2f^2/M_\phi^4$ to vary.
The shown bounds on $M_{t'}$ are the minimal values for an arbitrary 
value of $\lambda^2f^2/M_\phi^4$ within the [0,1/4] interval.
 For a two-parameter fit, $\Delta\chi^2=5.99$. The limit mentioned above
corresponds to the region near the origin in the left plot.

The bounds shown in Fig.~\ref{fig:littlest2} are quite
stringent. For all of the parameter space, the bounds on $M_{t'}$ exceed
$6$ TeV, introducing fine-tuning of more than 1\%. This is also
true for the parameter limit discussed in the previous paragraph, which
makes this limit seem uninteresting. However, as proposed in
Ref.~\cite{Chang:2003zn}, it is possible to enlarge one of the
$U(1)$'s to $SU(2)$ and make its coupling relatively strong so that
there exists an approximate custodial symmetry that suppresses the
coefficient $a_h$.

It is worth mentioning that LEP2 data included in our fit contributes
significantly to the constraints. For comparison, the plot on
the right in Fig.~\ref{fig:littlest2} shows the bounds from data
excluding LEP2 measurements for the $Y^f=Y_1^f$ case. The bounds are
significantly relaxed compared with the bounds obtained using all data. This
is because LEP2 experiments are very sensitive to the 4-fermion operators
$O_{ff'}^{s,t}$ generated by $Z'$ and $W'$ exchanges, while LEP1 and
other measurements are only sensitive to a few of them.

Other modifications of the model include gauging only $U(1)_Y$
\cite{Csaki:2003si} and  applying a $T$ parity \cite{Low:2004xc}. The
former is similar to the first limit discussed above,
where the coefficients $a_h$, $a_{hf}^s$, $a_{ff'}^s$ are suppressed
due to the lack of the $Z'$ boson contribution. The latter avoids generating
operators at tree level and thus constraints on the model from EWPTs
are less stringent.

\section{The $SU(6)/SP(6)$ model \cite{su6}}
\label{sec:su6}
The $SU(6)$ model has the same gauge structure as the $SU(5)$ model
but different global symmetry. The nonlinear sigma model terms have the same form
as in Eqs.~(\ref{kinetic}) and (\ref{derivative}). The  $\Sigma_0$
and $\Pi$ are six by six matrices:
\begin{eqnarray}
  \Sigma_0=\left(\begin{array}{cccc}&\bf{1}_2&&\\-\bf{1}_2&&&\\&&&1\\&&-1&\end{array}\right),
\quad
  \Pi=\frac{1}{\sqrt{2}}\left(\begin{array}{cc}{\bf{0_4}}&
     \begin{array}{cc}h_1&h_2\\-h^*_2&h_1^*\end{array}\\
      \begin{array}{cc}h^\dag_1&-h_2^T\\h_2^\dag&h_1^T\end{array}&{\bf{0_2}}
          \end{array}\right),
\end{eqnarray}
where the two Higgs doublets $h_1$, $h_2$ have the $SU(2)\times U(1)$ 
SM quantum numbers $({\bf2},+1/2)$ and $({\bf2},-1/2)$, respectively.
We have omitted the eaten fields from the $\Pi$ matrix, as well as omitted
singlet pseudo-Goldstone bosons that do not affect EWPTs.  We have
adopted the basis used in Ref.~\cite{Gregoire:2003kr} here, but changed
the definition of $h_2$ for convenience. The $[SU(2)\times U(1)]^2$
generators are also given in Ref.~\cite{Gregoire:2003kr}.

The gauge boson mixings are described by
Eqs.~(\ref{gaugemixing}) and (\ref{mixingangles}) and their  masses are
\begin{equation}
M_{W'}=\frac{g F}{2sc},\quad M_{Z'}=\frac{g' F}{\sqrt{8}s'c'}.
\end{equation}

Integrating out the heavy gauge bosons, we obtain
\begin{eqnarray}
a_h&=&-\frac{1}{F^2}[(c'^2-s'^2)^2+\frac12\cos^2(2\beta)],\nonumber\\
a_{hq}^t&=&a_{hl}^t=-\frac{1}{2F^2}(c^2-s^2)c^2,\nonumber\\
a_{hf}^s&=&\frac{2s'c'(c'^2-s'^2)}{F^2}\left(Y_2^f\frac{s'}{c'}-Y_1^f\frac{c'}{s'}\right),\nonumber\\
a_{lq}^t&=&a_{ll}^t=-\frac{c^4}{F^2},\nonumber\\
a_{ff'}^s&=&\frac{-8s'^2c'^2}{F^2}\left(Y_2^f\frac{s'}{c'}-Y_1^f\frac{c'}{s'}\right)
 \left(Y_2^{f'}\frac{s'}{c'}-Y_1^{f'}\frac{c'}{s'}\right).\label{su6ops}
\end{eqnarray}
Since only the vevs of $h_1$ and $h_2$ matter, we have combined
their contributions to a single $h$. The term proportional to $\cos^2(2\beta)$
in $a_h$ comes from expanding the kinetic term in Eq.~(\ref{kinetic}).

The Yukawa couplings can be constructed in several different ways.
Here we choose the Yukawas in a way similar to the $SU(5)$ model to
avoid extra corrections to EWPOs. For example, we add a pair of
fermions ($t'_L, t'_R$) and define
$\chi=(u_{3L},d_{3L},0,0,0,t'_L)^T$. Then the top Yukawa coupling
comes from the Lagrangian
\begin{equation}
\lambda_1
f\epsilon_{ijk}\epsilon_{xy}\overline{\chi_i}\Sigma_{jx}\Sigma_{ky}
u_{3R}+\lambda_2 f\overline {t'}_L t'_R,
\end{equation}
where $i,j,k\in\{1,2,6\}$ and $x,y\in\{3,4\}$. This Lagrangian only
induces mixing between the top quark and the heavy fermion and thus
does not affect EWPTs. 

\begin{figure}
\begin{center}
\includegraphics[width=\textwidth]{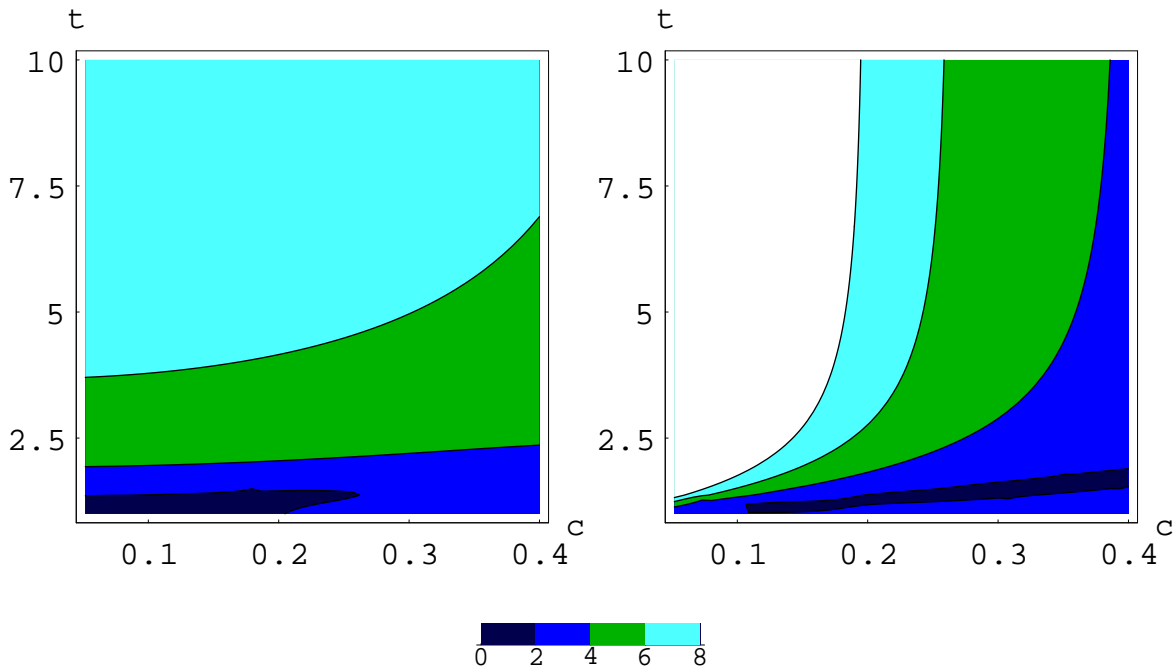}
\end{center}
\caption{Lower bounds in TeV on $M_{t'}$ (left) and $M_{W'}$ (right)
in the $SU(6)$ model as functions of $c$ and $t\equiv \tan\beta$ at 
95\% CL for $Y_1^f=Y_2^f$ and $s'=c'$. In the plots, $c\in [g/4\pi,0.4]$ and
$t \in[1,10]$.} \label{fig:su61}
\end{figure}

Comparing Eq.~(\ref{littlestops}) and Eq.~(\ref{su6ops}), we see
that the two sets of coefficients have similar structure. Thus we
expect that the constraints from EWPTs in the $SU(6)$ model to have
similar features as in the $SU(5)$ model. First, if the
fermions are charged under only one $U(1)$, the contributions
from the $Z'$ exchange are large and one expects tight
constraints. In this case, we have verified that the bounds on
$M_{t'}$ are greater than $6$~TeV for all choices of $c$, $c'$ and
$\tan\beta$. Second, the two limits that lead to suppression of
coefficients are present here as well. Instead of the triplet contribution
in the $SU(5)$ model, there is another custodial symmetry breaking
term generated by the nonlinear structure of the $\Sigma$ field
in this model. For comparison with the
$SU(5)$ model, when $s'=c'$ and $Y_1^f=Y_2^f$  we draw 95\%
CL bounds on $M_{t'}$ and $M_{W'}$ as functions
of $c$ and $\tan\beta$ in Fig.~\ref{fig:su61}. The ``near-oblique'' limit
discussed in Ref.~\cite{Gregoire:2003kr} corresponds to the region where
$\tan\beta\sim1$ and $c\sim0$ in Fig.~\ref{fig:su61}. In this limit,
the new particles could be so light that loop corrections have to be
considered to obtain accurate bounds, as has been done in
Ref.~\cite{Gregoire:2003kr}. Like in the $SU(5)$ model, the other
limit: $Y_1^f=Y^f$, $Y_2^f=0$ and $c,c'\ll1$ suppresses all
coefficients except $a_h$.

The two $U(1)$ generators can also be modified. Since the
global symmetry is larger than in the $SU(5)$ model, we have more
choices. For example, by redefining 
\begin{equation}
Y_1\rightarrow Y_1+bI+b'\mbox{Diag}\{{\bf1}_2,-{\bf1}_2,0,0\},\quad
Y_2\rightarrow Y_1-bI-b'\mbox{Diag}\{{\bf1}_2,-{\bf1}_2,0,0\},
\end{equation}
we can rescale the $Z'$ mass and the corresponding $a_i$. We can
also change the $U(1)$ generators in a way that the Higgs bosons are
charged differently under the two $U(1)$'s. For example, taking
\begin{equation}
Y_1=\left(\begin{array}{ccc}0_4&&\\&-\frac58&\\&&0\end{array}\right),\quad
Y_2=\left(\begin{array}{ccc}0_4&&\\&\frac18&\\&&\frac12\end{array}\right)
\end{equation}
yields the correct $U(1)_Y$ charges for $h_1$ and $h_2$, while the
coupling between $h_2$ and $Z'$ becomes
\begin{equation}
\frac12 igZ'^\mu(D_\mu h_2)^\dag(\frac{5c'}{8s'}-\frac{3s'}{8c'})h_2+h.c.
\end{equation}
There are similar terms for the $h_1$ to $Z'$ coupling.
Correspondingly, $a_{h}$ and $a_{hf}^s$ are no longer proportional
to $(c'^2-s'^2)$. Such changes affect the near-oblique condition, but
they do not introduce essentially new features in our analysis. It 
would be interesting if such structure comes naturally from some underlying theories.

\section{The $SU(3)\times U(1)$ models
\cite{Kaplan:2003uc,Skiba:2003yf,simplest}}
\label{sec:su3}

As effective theories, the little Higgs models do not need to be
anomaly free. The original $SU(3)$ model \cite{Kaplan:2003uc} is
anomalous, indicating that additional fermions must be present at the
cutoff. Anomaly-free versions of the model have also been
constructed \cite{simplest,Kong:2004cv}. However, they require
assigning different charges for different generations. Our methods
can not be applied to this generation-dependent model. Therefore,
 we concentrate on the original model only.

By integrating out the heavy fermions, we obtain the following
coefficients. The notation follows Ref.~\cite{simplest}. For one
generation,
\begin{eqnarray}
  a_{hl}^s&=&-a_{hl}^t=\frac14\frac{f_2^2}{F^2 f_1^2},\label{su3hlepton}\\
  a_{hq}^s&=&-a_{hq}^t=\frac14\frac{(\lambda_1^{u2}-\lambda_2^{u2})^2f_1^2f_2^2}
         {[(f_1\lambda_1^u)^2+(f_2\lambda_2^u)^2]^2 F^2}  \label{su3hquark0},
\end{eqnarray}
where $F^2=f_1^2+f_2^2$. For three generations, $\lambda^u_1$ and
$\lambda^u_2$ are flavor dependent and in general should be
$3\times3 $ matrices. However, in order to avoid large FCNCs, one
should set one of the matrices to be proportional to the identity matrix and
the other one to be hierarchical \cite{Kaplan:2003uc}. For example, if $\lambda^u_2$ is 
proportional to the identity matrix and $\lambda_1^u$ is hierarchical, for the
first two generations, $\lambda_1^u\ll\lambda_2^u$, so we can
approximate $\lambda_1^u\simeq0$. For the third generation, since
only the top quark mixes with the heavy quark the EWPTs are not
affected. The relation $a_{hq}^s=-a_{hq}^t$ is a consequence of the
fact that the bottom quark does not mix with the heavy fermion. We
can assign any values to $a_{hq}^s$ and $a_{hq}^t$ for the third
generation as long as the relation holds. Therefore we can ignore all terms containing
$\lambda_1^u$ in Eq.~(\ref{su3hquark0}). Similarly, if $\lambda_2^u$ is hierachical 
instead, we can ignore all terms containing $\lambda_2^u$. In this case the coefficients
$a_{hq}^s$ and $a_{hq}^t$ are identical to the coefficients of the corresponding 
lepton operators in  Eq.~(\ref{su3hlepton}) and the constraints on the model turn out to be
even less stringent.\footnote{We thank Heather Logan for drawing our attention to this possibility.}
Thus, we obtain the following flavor-independent operators 
\begin{eqnarray}
a_{hq}^s=-a_{hq}^t=\frac14\frac{f_1^2}{F^2 f_2^2} \quad 
    \mbox{ ($\lambda_1^u$ hierarchical)},\nonumber\\
a_{hq}^s=-a_{hq}^t=\frac14\frac{f_2^2}{F^2 f_1^2}  \quad
   \mbox{ ($\lambda_2^u$ hierarchical)}.\label{su3hquark}
\end{eqnarray}
By integrating out the heavy gauge bosons, we obtain
\begin{eqnarray}
a_h&=&-\frac{9}{4F^2}\frac{(1-\frac23x^2)^2}{(3+x^2)^2},\nonumber\\
a_{hf}^s&=&\frac{9}{4F^2}\frac{1-\frac23x^2}{(3+x^2)^2}(\sqrt{3}T^{8f}+x^2
Y_x^f),\nonumber\\
a_{ff'}^s&=&-\frac{9}{2F^2}\frac{1}{(3+x^2)^2}(\sqrt{3}T^{8f}+x^2
Y^f_x)(\sqrt{3}T^{8f'}+x^2Y^{f'}_x),\label{su3hgauge}
\end{eqnarray}
where $x=g_x/g$, while $T^{8f}=1/(2\sqrt{3}),1/(2\sqrt{3}),0,0,0$ and
$Y_x^f=1/3,-1/3,2/3,-1/3,-1$ are fermion charges for
$f=q,l,u,d,e$, respectively. Moreover, $g_x$ is related to $g$ and
$g'$ by
\begin{equation}
g_x^2=\frac{3g^2g'^2}{3g^2-g'^2}.
\end{equation}
The complete list of coefficients is given by 
Eqs.~(\ref{su3hlepton}), (\ref{su3hquark}) and (\ref{su3hgauge}).

In this model, the  mass of the  heavy fermion and the top Yukawa coupling
are given by~\cite{simplest}
\begin{equation}
M_{t'}^2=\lambda_1^2f_1^2+\lambda_2^2f_2^2,\quad
\lambda_t=\lambda_1\lambda_2\frac{F}{M_{t'}}.
\end{equation}
Unlike the $SU(5)$ and $SU(6)$ models, for given $f_1$ and $f_2$,
$M_{t'}$ is not uniquely determined because of the freedom in choosing the
ratios $\lambda_1/\lambda_2$ and $f_1/f_2$. Thus, $M_{t'}$ is not  tightly constrained.
However, the mass of the heavy gauge doublet $W'$ is
determined uniquely in terms of  $F$: $M^2_{W'}=g^2F^2/2$.

\begin{figure}
\begin{center}
   \includegraphics[width=\textwidth]{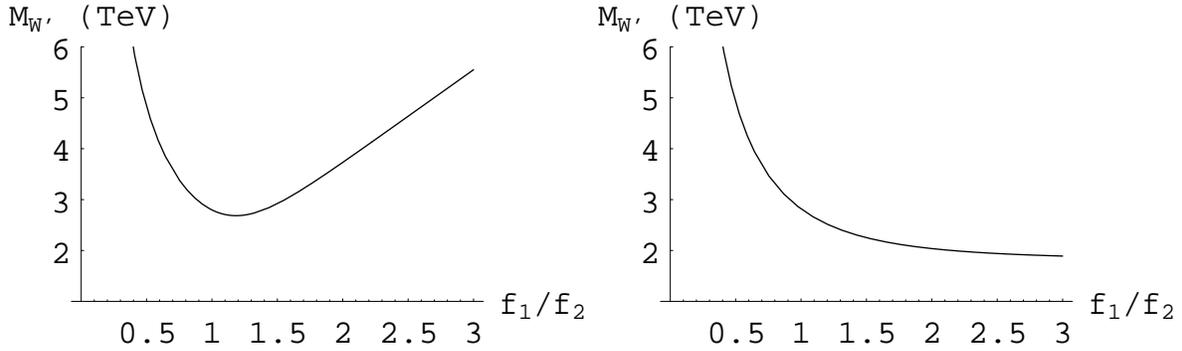}
\end{center}
  \caption{Lower bounds on $M_{W'}$ as a function of $f_1/f_2$ in the $SU(3)$ models. Left:
 $\lambda_1^u$ is hierarchical and $\lambda_2^u$ is proportional to identity matrix; right: 
 $\lambda_2^u$ is hierarchical and $\lambda_1^u$ is proportional to identity matrix. } \label{fig:su31}
\end{figure}

Given the ratio $f_1/f_2$, we can obtain the bound on $M_{W'}$.
Fig.~\ref{fig:su31} shows 95\% CL bounds on $M_{W'}$ as a function
of $f_1/f_2$. Since the contribution to the Higgs mass from $W'$ is
not significant until $M_{W'}>6$ TeV  the fine-tuning problem is not
severe. Depending on the choice of the Yukawa matrices in the quark sector corresponding to
 the two possibilities in Eq.~(\ref{su3hquark}), the least severe constraints are either
when $f_1/f_2\approx 1$ for  hierarchical $\lambda_1^u$, or when $f_1/f_2>2$
for hierarchical  $\lambda_2^u$.

The $SU(3)\times U(1)$ gauge sector can be embedded in different
global groups. The above discussion makes use of the setup of
Ref.~\cite{simplest}, where the coset space  is $[SU(3)/SU(2)]^2$.
But the constraints also apply to the model constructed in
Ref.~\cite{Skiba:2003yf}, where the coset space is $SU(9)/SU(8)$.
The model is also extended to an $[SU(4)/SU(3)]^4$ model with an
$SU(4)$ group gauged in Ref.~\cite{Kaplan:2003uc}. These variations
mainly stem from theoretical considerations, rather than the need to
avoid the experimental constraints.

\section{Summary and discussion}
\label{sec:conclusions}

Using effective field theory approach, we have obtained constraints from
EWPTs for three types of little Higgs models: $SU(5)/SO(5)$,
$SU(6)/SP(6)$ and $SU(3)\times U(1)$. We have carried out the
analysis by first integrating out the heavy fields to obtain the
effective operators, and then calculating the constraints from the
bounds on the linear combinations of the coefficients of these operators.

We gained two main benefits from following through this procedure.
First, expressing the corrections in terms of effective operators allowed us to treat the
corrections in a model-independent way, which we have done in
Ref.~\cite{Han:2004az}. Thus we can simply use the results in
Ref.~\cite{Han:2004az} once we acquire the coefficients of the
operators.  For the $SU(5)$ and $SU(6)$ models,
we use the heavy top mass $M_{t'}$ and the heavy triplet gauge boson
mass $M_{W'}$ to illustrate the bounds. We show that there exist
regions in the parameter space where the bounds on $M_{t'}$ and $M_{W'}$
are less than 2 TeV and 6 TeV respectively. In these regions, no
significant fine-tuning is required to yield light Higgs bosons. These
regions are in the vicinity of the points where the two $U(1)$
gauge sectors in the models have the same coupling strength and
fermions are charged identically under the two $U(1)$'s. Away from
this limit, the bounds are usually much tighter. In particular, for
the cases that fermions are charged under only one $U(1)$, we obtain
the bounds $M_{t'}>6$ TeV at 95\% CL for all of the parameter space,
requiring fine-tuning of more than 1\%. For the $SU(3)$ model,
$M_{t'}$ remains a free parameter for a given $F$. Thus we only
obtain the bounds for $M_{W'}$. At 95\% CL, $M_{W'}>1.8$ TeV. The
associated  fine-tuning is not significant in this case.

It is interesting that LEP2 results are very useful. 
The models induce many 4-fermion operators which LEP2
observables are sensitive to. The LEP2 results were not included in
most of previous papers considering EWPT constraints on littlest
Higgs models. As a comparison, we have checked our results for the
$SU(5)$ and $SU(6)$ models against
Ref.~\cite{Csaki:2002qg,Csaki:2003si}. We agree with the results in
the two references when we use the same set of observables, which
contain most precisely measured LEP1 and some low-energy observables.
After we include the LEP2 observables, the bounds become
significantly tighter.

In Ref.~\cite{Marandella:2005wd} the authors also obtain
constraints on little Higgs models from EWPTs including LEP2 data.
Technically, our approach is more general. The authors of
Ref.~\cite{Marandella:2005wd} have to assume that the models are
approximately universal and all corrections can be condensed in
four oblique parameters. In particular, the heavy gauge bosons
should couple to fermions ``universally'', which means that the
fermion currents that couple to the heavy gauge bosons are
proportional to the SM current. This is not true for the $SU(3)$
model, so the authors of Ref.~\cite{Marandella:2005wd}
have to neglect some of the EWPOs. In addition, they
do not discuss the effects of the heavy fermions in the $SU(3)$
model, which are not universal either. In the $SU(5)$ and $SU(6)$
models we have analyzed, the fermion charge assignments do yield
universal couplings between fermions and gauge bosons. Thus their
method apply as well. However, one could obtain non-universal
couplings from more general fermion assignments, which would make
their method inapplicable, but would not introduce any difficulty in
our approach. In \cite{Marandella:2005wd} only  the simplest fermion
assignments made in the original papers were considered, while we are 
interested in the variations which relax the electroweak constraints.

The second advantage of the effective theory approach is that it
makes it transparent how to  modify the models to avoid tight
constraints. From the compact but complete lists of operator
coefficients in the $SU(5)$ and $SU(6)$ models, we easily deduce  two
regions of parameter space that lead to suppressions of the corrections.
One of the regions for the $SU(6)$ model case is discussed and termed
``near-oblique'' limit in Ref.~\cite{Gregoire:2003kr}. In
Ref.~\cite{Chang:2003zn}, the other limit has been utilized in a
model with enlarged gauge sector and an approximate custodial symmetry.
With their weaker constraints, such model variations are interesting since 
they address the fine-tuning problem better than the original models.  From
experimental point of view, new particles are light enough to be
observable at the LHC only if there is no significant fine tuning.

\section*{Acknowledgments}
This research was supported in part by the US Department of Energy
under grant  DE-FG02-92ER-40704 and by Outstanding Junior Investigator grant.
WS thanks the Aspen Center for Physics for its hospitality.


\end{document}